\documentstyle[12pt]{article}
\def\a{\alpha}
\def\b{\beta}

\def\d{\delta}
\def\e{\epsilon}

\def\g{\gamma}

\def\k{\kappa}

\def\q{\theta}

\def\cu{{\cal U}}

\def\abs#1{\vert {#1} \vert}
\def\Hat#1{\widehat{#1}}                        % big hat
\def\frac#1#2{{\textstyle{#1\over\vphantom2\smash{\raise.20ex
        \hbox{$\scriptstyle{#2}$}}}}}                   % fraction
\def\su{\sum}
\newskip\humongous \humongous=0pt plus 1000pt minus 1000pt
\def\caja{\mathsurround=0pt}
\def\eqalign#1{\,\vcenter{\openup2\jot \caja
        \ialign{\strut \hfil$\displaystyle{##}$&$
        \displaystyle{{}##}$\hfil\crcr#1\crcr}}\,}
\newif\ifdtup

\parskip=\medskipamount                 % space between paragraphs (LaTeX)
\def\title#1#2#3#4{{\hbox to\hsize{#4 \hfill #3}}\par
        \begin{center} \vglue .5in {\large\bf #1}\\[.6in]
        {#2}\\[.1in]{Institute of Mathematical Sciences }\\{CIT Campus,
Madras 600 113, India}\\[1.5in]
        {\bf ABSTRACT}\\[.1in] \end{center} \begin{quotation}}  % title stuff
\def\Title#1#2#3#4#5#6#7#8{ {\hbox to\hsize{#8 \hfill UMDEPP #7\\} {
\hfill IMSc/ #6}}\par
        \begin{center} \vglue .4in {\large\bf #1}\\[.3in]
       {#2}\\[.1in]
        {\it {#3}}\\[.1in]
        {#4}\\[.1in] {\it {#5}}\\[.5in]
        {\bf ABSTRACT}\\[.1in]
        \end{center} \begin{quotation}}                 % " for 2 authors
\def\endtitle{\end{quotation}\newpage}                  % end title page

\def\sect#1{\bigskip\medskip \goodbreak \noindent{\bf {#1}} \nobreak \medskip}
\def\refs{\sect{REFERENCES} \footnotesize \frenchspacing \parskip=0pt}
\def\Item{\par\hang\textindent}

\begin{document}

\title{In\"on\"u - Wigner Contraction of Kac-Moody
Algebras}{Parthasarathi Majumdar}{IMSc/92-26}{June 1992}
{We discuss In\"on\"u-Wigner contractions of affine Kac-Moody
algebras. We show that
the Sugawara construction for the contracted affine algebra
exists only for a fixed value of the level $k$, which is determined
in terms of the dimension of the uncontracted part of the
starting Lie algebra, and the quadratic Casimir in the adjoint
representation. Further, we discuss contractions of $G/H$ coset
spaces, and obtain an affine {\it translation} algebra, which yields
a Virasoro algebra (via a GKO construction) with a central charge
given by $dim(G/H)$.}
\endtitle

\sect{1. Introduction}

Contractions of finite dimensional Lie algebras were introduced
several decades ago by In\"on\"u and Wigner$^1$, and applied
successfully to recover the Galilei group (and its
representations) from the Lorentz group. Subsequently, group
contractions were used to retrieve the Poincare group from the
de-Sitter group in various dimensions. This idea has been used very
successfully in obtaining irreducible representations of the
super-Poincare algebra from those of the super-de-Sitter
algebra which has a compact even (bosonic) subalgebra$^2$.

With the resurrection of string theory, affine Lie (Kac-Moody)
algebras have become the subject of very intense research$^{3,4}$.
Recently they have come into even more focus (especially in their
coset versions) following the formulation of the two dimensional
black hole as a gauged $SL(2,R)/U(1)$ Wess-Zumino-Witten
model$^5$. However, it is somewhat surprising that the issue of
contraction of affine Lie algebras has not received any attention
in the literature. Since the method of group contractions remains
a very powerful tool in obtaining noncompact (and non-semisimple)
algebraic structures from compact semisimple Lie algebras, it is
conceivable that the generalization of this to the infinite dimensional
case of affine Kac-Moody (KM) algebras will have ramifications for string
theories. With most attempts at construction of higher (i.e. $>$ 2)
dimensional noncritical strings stymied at present by the
so-called $c~=~1$ barrier this may be a worthwhile approach.

Indeed, if this barrier can be attributed to the existence of
tachyons in the spectrum, as has been argued most articulately by
Seiberg$^6$, then one might consider introducing spacetime
supersymmetry to eliminate them. Thus, noncritical Green-Schwarz
type theories, if they exist, might be a viable way out. From a
geometrical standpoint, such theories may be formulated as
Wess-Zumino-Witten models on (super)group manifolds corresponding
perhaps to the super-Poincare group (or some coset). Since, as
already mentioned, the super-Poincare algebra can be obtained
by contraction of superalgebras which are gradings of compact
semisimple Lie algebras, generalization of the
contraction procedure to the case of affine (super) Kac-Moody
algebras is likely to become important. This is the motivation of
this rather preliminary investigation.

Here we consider the simplest generalization of the In\"on\"u-Wigner
contraction applicable to affine Lie algebras (generically
denoted by ${\hat g}$ ) corresponding to finite dimensional
semisimple compact Lie groups (generically denoted by $G$). We
find rather stringent restrictions on the possible central
extensions of the contracted affine algebra, which can be best
understood in terms of the Cartan-Killing metric of $G$. A
straightforward Sugawara construction of the Virasoro algebra
associated with the contracted algebra seems to work only for a
fixed value of the level $k$ of the initial KM algebra,
which can be determined in terms of the dimension of the
uncontracted part of the algebra (assumed semisimple) and its
quadratic Casimir operator in the adjoint representation. These
restrictions are seen to disappear for affine $G/H$ coset spaces
under contraction. In this latter case, the Sugawara
construction (via the GKO$^4$ technique) is seen to exist
for any value of $k$ and leads to a Virasoro algebra with
central charge given by $dim(G/H)$. As we discuss briefly
in the concluding section, this could have applications
in non-critical superstring theories.

We should perhaps mention that our approach is not completely
rigorous, but we expect that the main results will survive a more
careful investigation.

\sect{2. In\"on\"u-Wigner Contraction}

Let ${T^a, a=1, \dots dimG}$ be a basis of generators of the Lie
algebra $g$ of the compact semisimple Lie group $G$. We make the
following decomposition of $g$ in the sense of a vector space:
$$ g~=~V_R~\oplus~V_C~~, \eqno(2.1)$$
where, $T^{\a}, \a=1, \dots dimV_R$ is a basis for $V_R$ and
$T^i, i=1, \dots dim V_C$ is a basis for $V_C$. We make the
following transformation on $T^a$
$$ {\tilde T}^a~=~\cu^{ab}(\e) T^b~~, \eqno(2.2)$$
where,
$$\eqalign{
\cu^{\a \b}~&=~\d^{\a\b}~~\cr
\cu^{ij}~&=~f(\e) \d^{ij}~~\cr
\cu^{i\a}~&=~0~=~\cu^{\a i}~~. } \eqno(2.3)  $$
Here $f(\e)$ has the property that $f(0)~=~0,~ f(1)~=~1$. Thus
the $T^{\a}$ are left invariant under this transformation. Now,
the Lie algebra $g$ is given by the commutation relations
$$[T^a,T^b]~=~if^{abc} T^c~~,\eqno(2.4)$$
so that under the decomposition (2.1) and the subsequent
transformation (2.2,3) the structure constants $f^{abc}$
transform into new ones denoted by ${\tilde f}^{abc}$, where
$$\eqalign {
{\tilde f}^{\a \b \g} &=~f^{\a\b\g}~~,~~{\tilde f}^{\a \b j}~ =~
f^{-1}(\e) f^{\a \b j}~~\cr
{\tilde f}^{\a i j} &=~ f(\e) f^{\a ij}~~,~~{\tilde f}^{ijk} ~=~
f^2(\e) f^{ijk}~~. } \eqno(2.5) $$
Thus if the limit $\e \rightarrow 0$ is to exist, one must
assume that the structure constants $f^{\a\b i}~=~0$. In this
limit therefore, one obtains, under this assumption,
a new algebra ${\tilde g}$, given by the commutation relations
$$\eqalign{
[{\tilde T}^{\a}, {\tilde T}^{\b}] &=~i f^{\a\b\g} {\tilde
T}^{\g}~\cr
[{\tilde T}^{\a}, {\tilde T}^i ] &=~i f^{\a ij} {\tilde T}^j \cr
[ {\tilde T}^i, {\tilde T}^j ] &=~ 0~~. } \eqno(2.6)  $$
The Lie algebra ${\tilde g}$ is the contraction of $g$ according
to In\"on\"u and Wigner. Observe that ${\tilde g}$ contains a
`translation' subalgebra of dimension $dim V_C$. However, one
should remark that contractions are not unique and one could
have certainly contracted $g$ leaving any other subalgebra
$V_R'$ invariant under the contraction. In the following, we
shall restrict this ambiguity to a certain extent by allowing
for $V_R$s which are themselves compact and semisimple.

It is of some interest to examine the effect on the
Cartan-Killing metric.
The Cartan-Killing metric $K_{ab}$ for $g$ can be expressed in
terms of the structure constants as $K_{ab}=f_{ac}^d f_{bd}^c$.
Thus, under the transformation (2.2,3) and in the limit $\e
\rightarrow 0$, the only non-vanishing components of the new
metric (i.e., that corresponding to ${\tilde g}$) are the ones
with Greek indices, and for these, ${\tilde K}_{\a\b}=K_{\a\b}$.
This property of the metric of the contracted algebra plays an
important role in restricting the possible central
extensions of the {\it affinized} contracted algebra, as we
shall show in the sequel.

\sect{3. Contraction of affine KM Algebras}

To generalize the foregoing to the case of affine KM algebras, we
first consider the classical case, i.e. the affine algebra
without the central extension (loop algebra). Following ref. 4, we
define maps $\g~:~S^1~\rightarrow~G$ which are given near the
identity by
$$ \g(z)~\approx ~ 1~-~i T^a \q^a(z)~, $$
where the parameters $\q^a(z)$ are holomorphic functions on the
unit circle $\abs{z}=1$, admitting the Laurent expansion
$$ \q^a(z)~~=~~\sum_n T^a z^n \q^a_{-n}~~. $$
Defining the affine basis $T^a_n \equiv T^a z^n$, we immediately
get the loop algebra ${\hat g}_0$
$$ [T^a_m, T^b_n]~~=~~if^{abc} T^c_{m+n}~~. \eqno(3.1) $$

It is now straightforward to proceed exactly as in the last
section. One decomposes the basis set of the generators
$$ {\hat g}_0~=~ {\hat V}_R \oplus {\hat V}_C~, $$
and applies the transformation  (2.2,3) to the corresponding
basis elements $T^{\a}_m$ and $T^i_m$. In the limit as $\e
\rightarrow 0$, we get the unique contraction ${\tilde {\hat
g}_0}$ of ${\hat g}_0$ given by
$$\eqalign{
[{\tilde T}^{\a}_m, {\tilde T}^{\b}_n] &=~ if^{\a \b \g}
{\tilde T}^{\g}_{m+n} \cr
[{\tilde T}^{\a}_, {\tilde T}^i ] &=~i f^{\a ij} {\tilde
T}^j_{m+n} \cr
[{\tilde T}^i, {\tilde T}^j ] &=~ 0~~. } \eqno(3.2) $$
The parameters ${\tilde \q}^a_{-n}$ are related to the original
$\q^a_{-n}$ through the relations
$$\q^{\a}_{-n}~=~{\tilde \q}^{\a}_{-n}~,~\q^i_{-n}~=~f(\e)
{\tilde \q}^i_{-n}~~. $$
Thus in the limit $\e \rightarrow 0$, the $\q^i_n$ vanish, quite
akin to the finite dimensional case.

We next incorporate the central extension for the affine KM
algebra which we call ${\hat g}$, given by
$$ [T^a_m, T^b_n]~=~if^{abc} T^c_{m+n}~+~\frac12 km \d^{ab}
\d_{m+n,0}~~. \eqno(3.3) $$
In order to generalize the contraction procedure described
above, one is faced with two choices. One could apply the
procedure directly to the affine KM algebra (3.3)
and thereby construct what one might call ${\tilde {\hat g}}$.
Alternatively, one might contract $g$ to ${\tilde g}$ first and
then affinize to obtain ${\hat{\tilde g}}$. The key issue
therefore is whether ${\tilde {\hat g}} \sim {\hat {\tilde g}}$ ?
Indeed, for a given subalgebra $V_R$ which is left invariant under
the contraction, the answer to the above query must be in the
affirmative if one is to give any meaning to the contraction
procedure in this infinite dimensional case.

The straightforward application of the contraction procedure to
(3.3) is seen to lead to the following commutation relations for
${\tilde {\hat g}}$ (we drop the tildes for ease of writing):
$$ \eqalign{
[T^{\a}_m, T^{\b}_n]& =~ if^{\a \b \g} T^{\g}_{m+n}~+~\frac12 km
\d^{\a\b} \d_{m+n,0}  \cr
[T^{\a}_m,T^i_n]&=~ if^{\a ij} T^j_{m+n} \cr
[T^i_m,T^j_n] &=~ 0. } \eqno(3.4)$$
Thus, one does not obtain an affine translation algebra, as one
might have expected. The central extension of ${\hat g}$ given
in terms of the level $k$ leads to a vanishing central extension
for the translation algebra upon contraction. If we were to
scale $k$ by a power of $f(\e)$, the first of the equations
(3.4) would be adversely affected as $\e \rightarrow 0$, which
is undesirable.

On the other hand, if we were to contract $g$ first and then
affinize in the standard fashion$^4$ (notwithstanding the
non-semisimple nature of ${\tilde g}$), na\"ively,
the translation subalgebra might be taken to have a structure
$$[T^i_m, T^j_n]~=~\frac12 {\tilde k}m \d^{ij} \d_{m+n,0}~.
\eqno(3.5) $$
But it is easy to rule out the central extension on the rhs of
(3.5) by using the Jacobi identity
$$ [T^{\a}_m,[T^i_n,T^j_p]]~+~cyclic~permutations~~=~~0~~.$$
Thus, given a contraction of $g$ which leaves a subalgebra $V_R$
invariant, there is a unique contraction given by (3.4)
independent of which of the choices in procedure one makes.

This uniqueness can be traced to the Cartan-Killing metric of the
contracted algebra. Recall that this is non-vanishing only in
directions left unaffected by contraction. It follows that free
field realizations of the contracted algebra exist only for the
compact, semisimple {\it uncontracted} subalgebra. E.g., for
contraction of ${\Hat {so(3)}}$ to ${\Hat {e(2)}}$, a free field
realization of the
latter can only be given for the ${\Hat {so(2)}}$ subalgebra.
Hence, the fact that one is able to retrieve a translation
algebra by this contraction seems not to be very useful from the
physical point of view, since all the physics appears to pertain
solely to the part of the original algebra that has been left
uncontracted.

These restrictions can be better articulated by considering a
Sugawara construction of the contracted algebra. The Virasoro
generators are defined as usual a la' Sugawara$^4$ as
$$\k L_n~\equiv~\sum_{m=-\infty}^{\infty}~ :~ T^a_{n-m} T^a_m
{}~:~, \eqno(3.6) $$
which, in terms of the generators $T^{\a}_m~,~T^i_m$ of the
contracted algebra can be rewritten as
$$\k L_n~=~\su_m ~:~\{T^{\a}_{n-m}T^{\a}_{m}~+~T^i_{n-m}T^i_m
\}~~. \eqno(3.7)$$
The constant $\k$ is to be determined by requiring that
$$\eqalign{
[L_n,T^{\a}_m] &=~-m T^{\a}_{m+n}~~\cr
[L_n,T^i_m] &=~ -m T^i_{m+n}~~. } \eqno(3.8) $$
Now, the first of these equations is easily seen to imply that
$$ \k~~=~~k~+~c_{V_R}~~, \eqno(3.9a) $$
where $c_{V_R}$ is the quadratic Casimir operator of the
uncontracted subalgebra $V_R$ in the adjoint representation. The
second of the equations (3.8) leads to a strange result
$$ \k~~=~~{F \over {2 dim V_C}}~~, \eqno(3.9b) $$
where, $F~\equiv~f^{\a ij} f^{\a ij} $. One is thus led to the
very important restriction that the Sugawara construction,
applied straightforwardly, works only for a specific value of
the level $k$ (obtained by solving (3.9) for $k$).

Indeed, for
the contraction of ${\Hat {so(3)}}$ to ${\Hat {e(2)}}$, this value
of k is easily seen to be unity. The Sugawara construction would
then imply that the central charge of the Virasoro algebra is
${\tilde c}=1$, a result that is identical in this case to the
Virasoro central charge for ${\Hat {so(3)}}$. As is well-known,
the level 1 ${\Hat {so(3)}}$ can be realized in terms of a single boson.
Thus the contracted algebra in this case also has a free boson
realization. This corroborates our earlier
remarks regarding possible free field realizations of the
contracted affine algebra.

The Virasoro algebra ensues after a straightforward calculation
using eqn.s (3.5-6) and yields a central charge ${\tilde c}_k$
given by
$${\tilde c}_k~~=~~{k dimV_R \over {\k}}~~. \eqno(3.7)$$
Clearly, for the contraction of ${\Hat {so(3)}}$ to ${\Hat
{e(2)}}$, we have, with $k=1$, ${\tilde c}_k=1$. This therefore
implies that our contraction procedure leads uniquely to the
realization of the contracted algebra in terms of a single free
massless scalar field. In this particular case of $so(3)$, the
central charge for the level one KM algebra is the same as that
of the contracted algebra $e(2)$. This is not generically true.
But the above restriction does have the disconcerting feature
that the translation subalgebra (obtained upon contraction)
is rendered physically irrelevant. The contraction procedure is
then devoid of any physical interest, at least in its simplest
version. It is conceivable that a more complicated contraction,
e.g., one in which the contraction parameter $\e$ is itself
taken to be a function $\e(z)$ on the circle
$S^1$, will yield a more physically interesting result. We do not
pursue such alternatives in this short note. Instead we analyze
contractions of affine coset spaces, generalizing the standard
procedure appropriate to finite dimensional $G/H$ cosets$^7$.

\newpage
\sect{4. Contraction of Coset Spaces}

Referring back to our discussion in Section 2, let us identify
the subalgebra $h$ corresponding to the subgroup $H$ with respect
to which $G$ is to be quotiented, with the subalgebra $V_R$ which
is left undisturbed by the contraction. In that case, the
relevant commutation relation is
$$[T^i,T^j]~=~i(f^{ij \a} T^{\a}~+~f^{ijk} T^{k} )~~.\eqno(4.1)$$
Proceeding as in Section 2, one obtains in the limit $\e
\rightarrow 0$, a translation algebra of dimensionality $dim(G/H)$
$$ [ {\tilde T}^i, {\tilde T}^j ]~~=~~0~~, \eqno(4.2)$$
where we have put back the tildes to exhibit that these
correspond to the contracted generators. One can then affinize
(4.2) in the standard fashion and obtain an affine translation
algebra with a central extension
$$ [{\tilde T}^i_m, {\tilde T}^j_n]~~=~~\frac12 {\tilde k}
\d^{ij} \d_{m+n,0}~, \eqno(4.3)$$
where ${\tilde k} \neq k$ in general.

Before discussing the Sugawara construction for (4.2), let us
mention that (4.3) is the unique (in the aforementioned sense)
contraction of the affine coset
space for $G/H$. In other words, if we affinize (4.1) first to
get
$$ [T^i_m, T^j_n]~~=~~i f^{ij a} T^{a}~+~ \frac12 k \d^{ij}
\d_{m+n,0} ~~\eqno(4.4)$$
and then contract (4.4) as in section 3, we do get back (4.3)
provided we rescale the level $k$ appropriately to ${\tilde k}$.
Since the central extension given by $k$ is an operator which
commutes with all the generators $T^i$, it is legitimate in
the contraction procedure
to transform it as well. This was ruled out
in the last section because there such a transformation would
have affected the central extension corresponding to the
uncontracted part of the algebra, a feature that we wanted to avoid.

For the Sugawara construction appropriate to the contracted
affine coset algebra (4.3), we follow ref. 4 and define the
Virasoro generators as
$$ {\tilde \k} {\tilde L}_n~\equiv~\su_m ~:~{\tilde T}^i_m {\tilde
T}^i_{n-m}~:~.\eqno(4.5)$$
Requiring as before that
$$[{\tilde L}_n, {\tilde T}^i_m ]~=~-m {\tilde T}^i_{m+n}~\eqno(4.6)$$
immediately fixes the constant ${\tilde \k}={\tilde k}$. The
central charge of the corresponding Virasoro algebra is then easily
calculated by evaluating the $[{\tilde L}_n, {\tilde L}_m]$ commutator
explicitly,
and one obtains ${\tilde c}~=~dim(G/H)$, a result that one has
already anticipated given that (4.3) is an affine translation
algebra in a space of dimensionality $dim(G/H)$. Notice that the
central charge is independent of the level ${\tilde k}$.

Thus the contracted affine $G/H$ coset space can be given a free
field realization by $dim(G/H)$ massless scalar fields in two
dimensions, corresponding to a free bosonic string in these
dimensions. If $dim(G/H) \neq 26$, such a theory will be
consistent only when the Liouville mode is taken into account.

\sect{5. Discussion}

We have not yet analyzed the consequences of the results of the
last section for general gauged WZW models. Presumably, the
contraction procedure collapses the interacting WZW theory into a
free field theory. What this means for the two dimensional black
hole formulated as a gauged $SL(2,R)/U(1)$ WZW model is an
interesting question. A general theory of contraction of
irreducible representations and field theoretic realizations of
KM algebras is indeed in order, but beyond the scope of this
short note.

Another interesting application of the foregoing analysis could
be to the contraction of affine {\it super}-Lie algebras and
super-cosets, especially the case of the quotient of a supergroup
by a bosonic subgroup. Henneaux and Mezincescu$^8$ have given a
formulation of the type II ten dimensional Green Schwarz
superstring as a WZW model on a super-coset
$$ {{N=2~ D=10~ super-Poincare~ group} \over {SO(9,1)}}~~. $$
For the $N=1$ case, similar formulations have also been
considered, via reduction from super-Chern Simons theories
wherein the gauge fields take values in the appropriate
super-coset$^9$. The question now is, whether Green Schwarz
superstrings in lower dimensions (3,4, e.g.) can be considered as
gauged WZW models on super-cosets. If so, then it might be useful
to begin with some supergroup like $SU(2|1)$ which is a grading of a
classical group, and then contract the quotient $SU(2|1)/SO(3)$
to get
$${{N=1~ D=3~ Euclidean~ Poincare~ superalgebra} \over {SO(3)}}~.
$$
A WZW model on such a coset might be a candidate for a
non-critical `Green Schwarz string' in three dimensional
Euclidean superspace.  We hope to
report on these and other related issues in a future publication.

I would like to thank H. Riggs, T. Jayaraman, H. Schnitzer and
especially B. Rai for helpful discussions. This work was completed
during a visit to the International Centre for
Theoretical Physics, Trieste, Italy, and I thank the Centre for
its kind hospitality.

\refs

\Item{[1]} In\"on\"u and E. Wigner, Proc. Natl. Acad. Sci. USA
{\bf 39}, 510 (1956). R. Gilmore, {\it Lie Groups. Lie Algebras
and their Applications}, Wiley, New York (1974).
\Item{[2]} See, e.g., P. Freund {\it Introduction to
Supersymmetry}, Cambridge University Press (1986) and references therein.
\Item{[3]} V. Kac, {\it Infinite Dimensional Lie Algebras - an
Introduction}, Birkh\"auser, Boston (1983).
\Item{[4]} P. Goddard and D. Olive, Int. Jou. Mod. Phys. {\bf
A1}, 303 (1986).
\Item{[5]} E. Witten, Phys. Rev. {\bf D44}, 321 (1991).
\Item{[6]} N. Seiberg, {\it Notes on Liouville Theory and 2d
gravity}, Rutgers preprint (1990).
\Item{[7]} R. Gilmore, ref. 1.
\Item{[8]} M. Henneaux and L. Mezincescu, Phys. Lett. {\bf 152
B}, 340 (1985).
\Item{[9]} M. Green, Phys. Lett. {\bf B 223}, 157 (1989); P.
Majumdar, Phys. Rev. {\bf D45}, 2883 (1992).

\end{document}